\pdfoutput=1
\documentclass[letterpaper,twocolumn]{aastex6}

\usepackage{graphicx}
\usepackage{natbib}
\usepackage{amsmath}
\usepackage{url}
\newcommand       \be		{\begin{equation}}
\newcommand       \ee		{\end{equation}}

\begin{document}

\title{Magnetospheric Truncation, Tidal Inspiral, \\and the Creation of Short and Ultra-Short Period Planets}

\author{Eve J. Lee\altaffilmark{1}, Eugene Chiang\altaffilmark{1,2}}
\altaffiltext{1}{Department of Astronomy, University of California, Berkeley, CA 94720-3411, USA; evelee@berkeley.edu}
\altaffiltext{2}{Department of Earth and Planetary Science, University of California Berkeley, Berkeley, CA 94720-4767, USA}

\begin{abstract}
Sub-Neptunes around FGKM dwarfs are evenly distributed 
in log orbital period down to $\sim$10 days,
but dwindle in number at shorter periods.
Both the break at $\sim$10 days
and the slope of the occurrence rate 
down to $\sim$1 day can be  
attributed to the truncation of protoplanetary disks
by their host star magnetospheres at co-rotation.
We demonstrate this by deriving planet occurrence
rate profiles from empirical distributions of pre-main-sequence
stellar rotation periods.
Observed profiles are
better reproduced when planets 
are distributed randomly in disks---as might be
expected if planets formed in situ---rather
than piled up near disk edges, as would
be the case if they migrated in by disk torques.
Planets can be brought from disk edges to
ultra-short ($< 1$ day)
periods by asynchronous equilibrium
tides raised on their stars.
Tidal migration can account for how
ultra-short period planets (USPs) are more 
widely spaced than their longer period counterparts.
Our picture provides a starting point for understanding 
why the sub-Neptune population drops at $\sim$10 days 
regardless of whether the host star is of type 
FGK or early M.
We predict planet occurrence rates around A stars
to also break at short periods, but at $\sim$1 day
instead of $\sim$10 days because A stars rotate
faster than lower mass stars
(this prediction presumes that
the planetesimal building blocks of planets
can drift inside the dust sublimation radius).
\end{abstract}

\section{Introduction}
\label{sec:intro}
 
The {\it Kepler} mission,
in combination with radial velocity surveys,
has revealed that sub-Neptunes, with radii $R < 4 R_\oplus$, 
are fairly ubiquitous at orbital periods
$P \lesssim 100$ days \citep[e.g.,][]{fressin13,dressing15}.
Microlensing indicates that they occur just as frequently beyond, at least around M stars
\citep[e.g.,][]{clanton15}.

\begin{figure}[!h]
    \centering
    \includegraphics[width=0.5\textwidth]{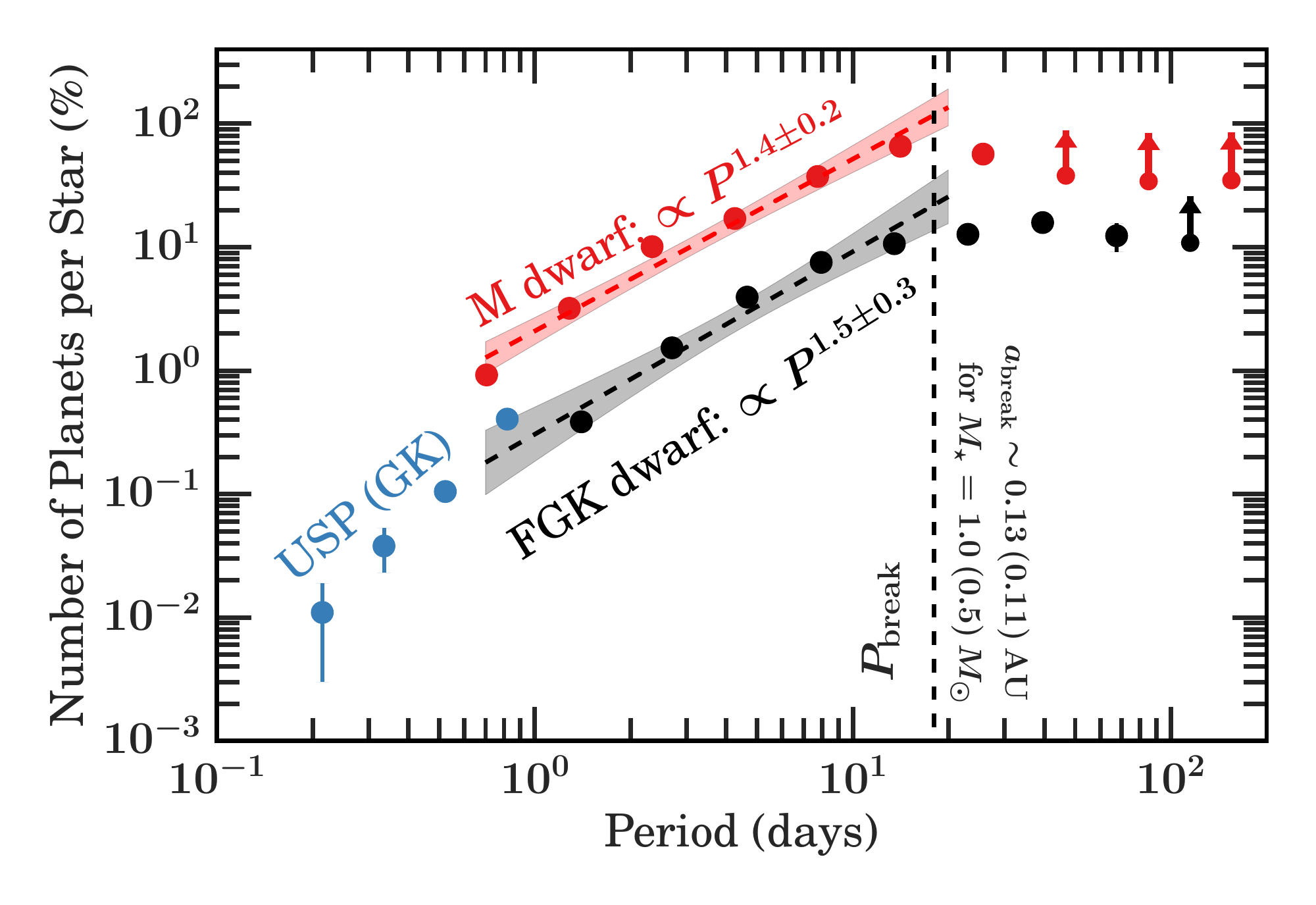}
    \caption{Occurrence rates of sub-Neptunes 
    ($R < 4 R_\oplus$) 
    orbiting FGK dwarfs \citep{fressin13}
    and early M dwarfs \citep{dressing15}. A distinctive
    break at $P_{\rm break} \sim 10$ days divides
    long-period planets at $P > P_{\rm break}$ from
    their less common, short-period
    counterparts
    at $P < P_{\rm break}$.
    Strikingly, short-period planets around FGKM host stars all 
    appear to be distributed according to
    $dN /d\log P \propto P^{1.4-1.5}$.
    We also distinguish ``ultra-short period'' (USP) planets 
    at $P < 1$ day, using data from \citet{sanchis-ojeda14}.
    All planets 
    are labelled according to their host star
    spectral types.
    Points without arrows correspond
    to sub-Neptunes larger than $0.5 R_\oplus$
    for M dwarf hosts, and larger than
    $0.8 R_\oplus$ for FGK hosts.
    Points with arrows represent 
    sub-Neptunes larger than $1 R_\oplus$ (M dwarfs),
    and larger than $1.25 R_\oplus$
    (FGK dwarfs).
    }
    \label{fig1}
\end{figure}

Though commonplace overall, sub-Neptunes (a population
including Earths and super-Earths)
are more likely to be found at some orbital periods than others.
At longer periods, they appear more-or-less evenly distributed
across logarithmic intervals in $P$.
But at shorter periods, sub-Neptunes are less common.
The occurrence rate as a function of orbital period
follows a broken power law, with a break at
$P_{\rm break} \sim 10$ days \citep[e.g.,][]{youdin11,mulders15}.
Inside $\sim$10 days, the occurrence rate scales approximately 
as $dN/d\log P \propto P^{1.5}$, while
beyond $\sim$10 days, the occurrence rate plateaus.
Remarkably, as we show in Figure \ref{fig1},
these power law slopes characterize planets found around stars
having widely varying spectral types, from FGK \citep{fressin13}
to early M \citep{dressing15}.\footnote{The Trappist-1 planetary
system \citep[][and references therein]{gillon17} is hosted by a late M dwarf.}

\begin{figure}
    \centering
    \includegraphics[width=0.5\textwidth]{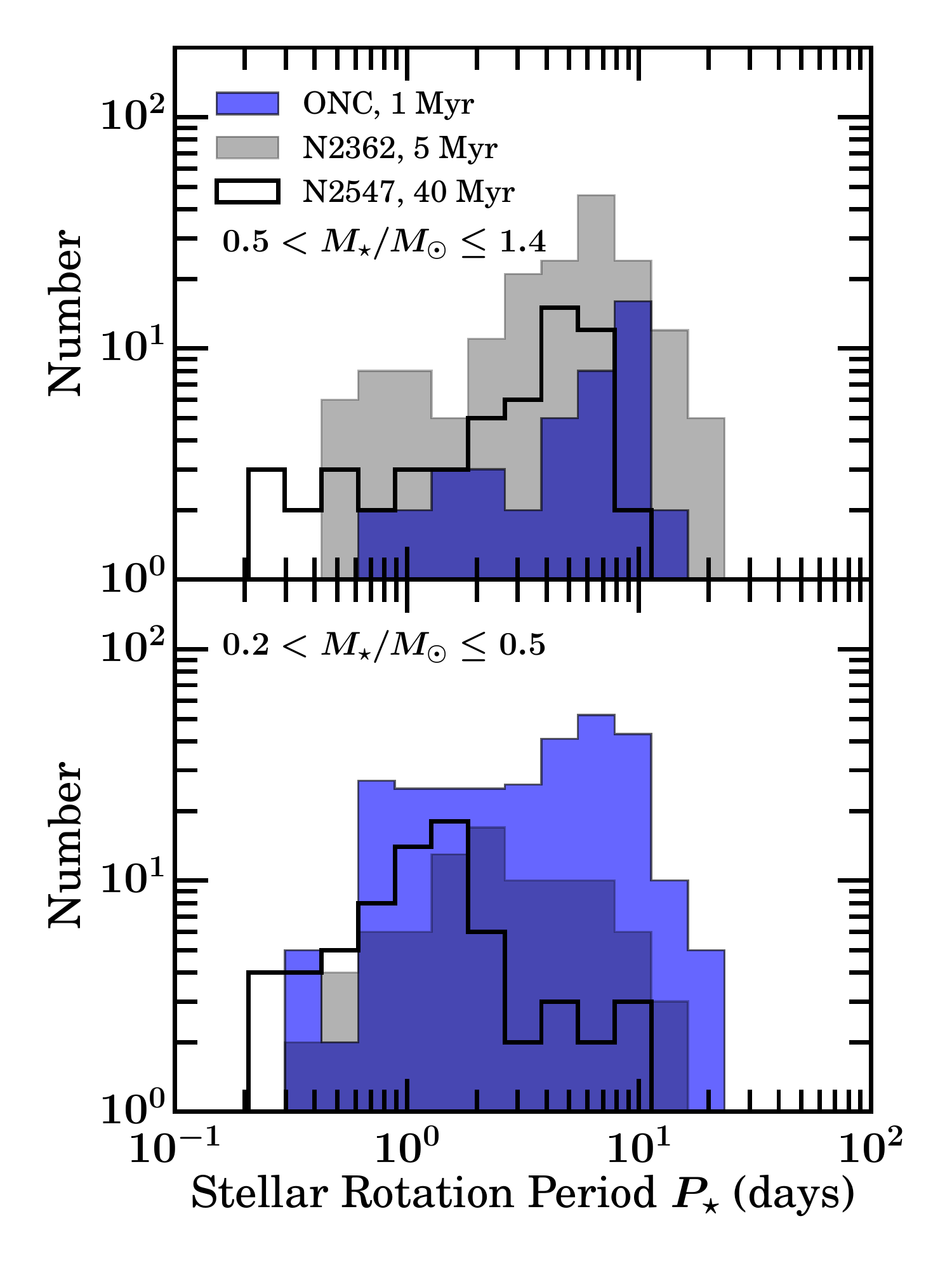}
    \caption{Histograms of pre-main-sequence stellar rotation periods $P_\star$ in clusters of various ages: the Orion Nebula Cluster (ONC, \citealt{herbst02}, \citealt{rl09}), NGC 2362 \citep{irwin08}, and NGC 2547 \citep{irwin08b}.
    Rotation periods are measured from 
    periodic light curve variations due to starspots.
    For higher mass stars (top panel), the distribution of $P_\star$ for all three clusters peaks at $\sim$10 days and falls off at shorter periods. Lower mass stars (bottom panel) tend to rotate faster at the same age.
    Older clusters harbor more rapidly rotating stars,
    as expected from stellar contraction. 
    See \S\ref{ssec:stellarmass}
    for a discussion of these trends.}
    \label{fig2}
\end{figure}

The lower occurrence rate of planets
at $P < P_{\rm break}$ (hereafter ``short-period'' planets)
may reflect truncation of their parent disks---perhaps
by the magnetospheres of their host
stars (e.g., \citealt{mulders15}; see also \citealt{plavchan13}).
The theory of ``disk locking'' posits
that the inner disk edge co-rotates with the host 
star in equilibrium 
\citep[e.g.,][]{ghosh79,konigl91,ostriker95,long05,romanova16}. 
Disk locking is supported observationally 
by ``dippers,'' young low-mass stars
with relatively evolved disks that exhibit material
lifted out of the disk midplane, presumably
by magnetic torques, near the co-rotation radius 
\citep{stauffer15,ansdell16,bodman16}.\footnote{Stellar 
magnetic fields measured
from Zeeman broadening do not
correlate well with field strengths predicted from magnetospheric
truncation (see the review by \citealt{bouvier07} and references
therein). However, relaxing the assumption that the stellar field
is a pure dipole brings theory into closer agreement with observations  \citep[see, e.g.,][and references therein]{cauley12}.}
Rotation periods $P_\star$ of stars 1--40 Myr old, displayed
in Figure \ref{fig2}, range from 0.2 to 20 days, squarely
in the range occupied by short-period planets.
Moreover, the distribution of $P_\star$ peaks near $\sim$10 days
and falls toward shorter periods, mirroring, at least
qualitatively, the decline in planet occurrence rate at
$P < P_{\rm break}$.\footnote{ All stellar rotation periods $P_\star$ shown in Figure \ref{fig2} are measured from
periodic light curve variations driven by starspots.
Photometric retrieval of $P_\star$ is biased against
young and especially
active T Tauri stars exhibiting irregular variability
caused by strong and unsteady disk accretion.
This bias is small---\citet{herbst02} estimate that 
their sample of periodic stars is incomplete
by 8--15\%---and its effect on our analysis should be
still more muted because we take
the orbital architecture of planets to be established
during the latest stages of disk evolution, when the disk
has largely (but not completely) dissipated
(see Section \ref{step2}).
As for the clusters NGC 2362 and NGC 2547,
\citet{irwin08} and \citet{irwin08b} find, respectively, that
their 
surveys are complete down to $\sim$0.2$M_\odot$ and across
all measured rotation periods.}
In this paper we develop, in quantitative
detail, this possible connection between
disk truncation at co-rotation and the occurrence rate profile
of short-period planets.

Planets that form outside the inner disk truncation radius
can be brought to the disk
edge, and even transported inside it,
by a variety of migration mechanisms.
These include disk torques (e.g., migration types I and II;
see the review by \citealt{kley12}),
stellar tidal friction, and planet-planet gravitational
interactions. We model the first two processes
in this paper but not the third. Below we briefly review
migration by planet-planet interactions and
explain why it cannot produce
the bulk of short-period sub-Neptunes:

\begin{enumerate}

\item Secular excitation of eccentricities in a
multi-planet system, coupled with
tidal circularization, can shrink planetary orbits
and produce short-period planets
\citep[e.g.,][]{wu11,hansen_murray15,matsakos16}. 
The sub-Neptune 55 Cnc e may have been so nudged
onto its current 0.7-day orbit \citep{hansen_zink15}.
But general-relativistic precession at short periods
usually defeats secular forcing by perturbers less
massive than Jupiter.
While many short-period sub-Neptunes are members
of multi-planet systems
\citep[e.g.,][and references therein]{steffen16,pu15}, 
their nearby companions are
not typically gas giants and therefore cannot compete
against general-relativistic precession.
55 Cnc e is atypical because its four neighbors
range in mass from 0.2 to 5 $M_{\rm J}$
(where $M_{\rm J}$ is the mass of Jupiter),
creating a web of strong secular resonances.

\item Violent planet-planet scatterings (i.e., close
encounters) can deliver planets from large
stellocentric distances to small ones
where their orbits can be circularized by tides.
But tidal circularization takes time, on the order
of Myrs to Gyrs for periapse distances
between 0.03 to 0.1 AU. During that time,
planets may collide with others and fail to attain short periods.
When bodies are on eccentric and crossing orbits,
encounters/mergers are more likely to occur at
large rather than small distances because bodies spend less time
near their periastra, and because interactions are less
gravitationally focussed at small distances.
High-eccentricity migration of short-period 
sub-Neptunes is hard to reconcile with the observation
that such planets are commonly found with 
close exterior neighbors.\footnote{By contrast,
high-eccentricity migration of hot Jupiters
is compatible with the observation that they have
distant, not close, exterior companions \citep{bryan16}.}

\end{enumerate}

In this work we attempt to explain the observed
occurrence rate profiles of short-period
planets, focusing on reproducing the break period $P_{\rm break}$
and the shape of the decline in planet frequency
at $P < P_{\rm break}$.
We take as our starting point circumstellar disks
whose inner edges are located at co-rotation
with their host stars, drawing disk truncation
periods directly from observed stellar rotation
periods (i.e., Figure \ref{fig2}). Outside
the disk truncation radius, we consider 
separately the possibility that sub-Neptunes migrate inward
by torques exerted by residual disk gas, and the possibility
that disk-driven migration is negligible,
i.e., that planets form in situ.
In either case, we account for orbital decay 
driven by asynchronous equilibrium tides
raised by planets on their host stars. 
We examine whether tidal friction can 
create ``ultra-short period'' planets
(USPs with $P < 1$ day) from short-period planets, in proportions
similar to those observed.
Our Monte Carlo model incorporating these various 
ingredients is presented in Section \ref{sec2}.
We summarize in Section \ref{sec:discuss},
highlighting those areas 
that need the most shoring up, and venturing a
few predictions.

\vspace{0.2cm}

\section{Monte Carlo Simulations}
\label{sec2}

We construct a Monte Carlo (MC) model
for the occurrence rates of sub-Neptunes
vs.~orbital period. Our goal is to identify
those parameters related to disk truncation,
disk-driven planet migration, and tidal 
inspiral that can reproduce
the shapes of the observed
occurrence rate profiles as shown in Figure \ref{fig1}.

The following subsections (\S\ref{step1}--\ref{step3})
detail the three steps taken to construct a given MC model.

\subsection{Step 1: Draw Disk Truncation Periods \\from Stellar Rotation Periods}\label{step1}

We create $N_{\rm disk} = 20000$ disks,
each truncated at its own innermost edge at
period $P_{\rm in}$.
The disks are assumed to be magnetospherically
truncated at co-rotation
with their host stars (i.e., we assume disk locking):
each value of $P_{\rm in}$ is drawn
directly from an empirical distribution
of stellar rotation periods $P_\star$.
A given MC model utilizes rotation periods measured for
one of three
stellar clusters: the Orion Nebula Cluster (ONC, age of 1 Myr),
NGC 2362 (5 Myr), and NGC 2547 (40 Myr).
For MC models of planets around
FGK dwarfs, we draw $P_{\rm in} = P_\star$ using
stars with mass $0.5 M_\odot < M_\star \leq 1.4 M_\odot$
(upper panel of Figure \ref{fig2}).
For MC models of planets around M stars,
we do the same for stars of mass
$0.2 M_\odot < M_\star \leq 0.5 M_\odot$
(lower panel of Figure \ref{fig2}).

\subsection{Step 2: Lay Down Planets Following Either\\``In Situ'' Formation or ``Disk Migration''}
\label{step2}

Each disk is taken to contain $N_{\rm planet} = 5$ planets.
Their initial periods---initial as in
before stellar tides have acted---are
decided according to one of two schemes.
``In-situ'' formation models are very simple:
the pre-tide planet
periods are distributed randomly and uniformly
in log period from $P_{\rm in}$ to 
$P_{\rm out} = 400$ days.
The logarithmic spacing befits planetary systems that form by
in-situ, oligarchic accretion, whereby planets are separated
by a multiple number of Hill radii \citep[e.g.,][]{kokubo12}.
Our specific choices of $N_{\rm planet} = 5$ 
and $P_{\rm out} = 400$ days affect only
the normalization 
of the occurrence rate profile; this normalization
is adjusted later to fit the observations
(see captions to Figures \ref{fig4} 
through \ref{fig8}).
The aim of this paper is to reproduce the breaks and 
slopes of the observed occurrence rates,
not their absolute normalizations.

The alternative ``disk migration''
models start the same way as in-situ models,
but transport planets to shorter pre-tide periods according
to the Type I migration rate:
\begin{equation}
    \dot{a} = \frac{2T_{\rm L}}{M_{\rm p}\Omega a}
    \label{eq1}
\end{equation}
where $a$ is a given planet's orbital radius,
$M_{\rm p}$ its mass, $\Omega$ its
orbital angular frequency, 
and $T_{\rm L}$ the combined Lindblad and co-rotation torque
exerted by the disk on the planet \citep[e.g.,][]{kley12}:
\begin{equation}
    T_{\rm L} = -\xi \left(\frac{M_{\rm p}}{M_\star}\right)^2 \left(\frac{a}{h}\right)^2 \Sigma_{\rm gas} a^4 \Omega^2
    \label{eq2} \,,
\end{equation}
where $\xi = (1.36 + 0.62 \beta + 0.43 \gamma)$ is an order-unity
constant calibrated from simulations \citep{dangelo10}, and 
$\beta$ and $\gamma$ characterize an assumed power-law
gas disk of surface density
\begin{equation}
\Sigma_{\rm gas} = \Sigma_{\rm gas,0} (a / a_0)^{-\beta} \exp(-t/t_{\rm disk}) 
\end{equation}
and temperature
\begin{equation}
T = T_0 (a / a_0)^{-\gamma} \,.
\end{equation}
Other variables include the disk scale height
$h = \sqrt{kT/\mu m_{\rm H}}/\Omega$, the gas
mean molecular weight $\mu$, 
the mass of the hydrogen atom $m_{\rm H}$, and
Boltzmann's constant $k$.

For simplicity, regardless of whether the host stars
modeled are FGK dwarfs or M dwarfs,
we fix $a_0 = 0.1$ AU,
$T_0 = 1000$ K, $\gamma = 1/2$, 
$\mu = 2$ and $M_{\rm p} = 5 M_\oplus$.
We fix $M_\star = 1 M_\odot$ for FGK dwarfs 
and $M_\star = 0.5 M_\odot$ for M dwarfs.
The errors introduced by fixing these variables
are subsumed to some extent by large uncertainties in the
disk parameters $\beta$, $\Sigma_{\rm gas,0}$,
and the dissipation timescale $t_{\rm disk}$. 
We now describe and comment on our choices
for these three parameters:

\begin{enumerate}
\item For a given MC model, we set $\beta$
to one of three values: $\beta \in \{1,2,3\}$.
If we assume that a disk's gas traces its solids
(i.e., a spatially constant gas-to-solids abundance 
ratio), then these $\beta$-values encompass the
range of quoted slopes in the literature,
as inferred from
observations of exoplanets comprising
mostly solids:
$\Sigma_{\rm solid} \propto a^{-1.5}$ to $a^{-2.4}$
\citep[e.g.,][]{hansen_murray12,chiang13,schlichting14,hansen15}. 
Note that
$\Sigma_{\rm solid} \sim dN /d\log P \times M_{\rm p} / a^2 \propto a^{-2}$
follows immediately from $dN/d\log P \propto P^0$
(this flat slope is seen at long periods in Figure \ref{fig1})
if the characteristic planet mass $M_{\rm p}$
does not vary much with distance.

\item We experiment with two normalizations for the gas surface
density at $a_0 = 0.1$ AU:
$\Sigma_{\rm gas,0} \in \{ 40, 400 \}$ g cm$^{-2}$.
These choices yield rather gas-poor disks, depleted
by 3--4 orders of magnitude relative to solar-composition
``minimum-mass nebulae''. Such highly gas-depleted
environments are motivated
by the need to defeat gas dynamical friction before
sub-Neptune cores can form by the mergers of protocores
\citep[][their Figure 5]{paper3}. 
We assume that Type I migration proceeds normally in
these gas-poor disks and discuss this assumption
in Appendix \ref{sec:app1}.

\item For the gas exponential decay timescale,
we try $t_{\rm disk} \in \{ 1, 10 \}$ Myr, which overlaps with
the range of typical disk lifetimes reported from
observations \citep{mamajek09,alexander14,pfalzner14}.
\end{enumerate} 

Having assigned all parameters for our disk migration model,
we then migrate every planet inward by integrating 
equation (\ref{eq1}) over a time interval of 
$10\times t_{\rm disk}$ (10 e-foldings of the disk
gas density).

\begin{figure}
    \centering
    \includegraphics[width=0.5\textwidth]{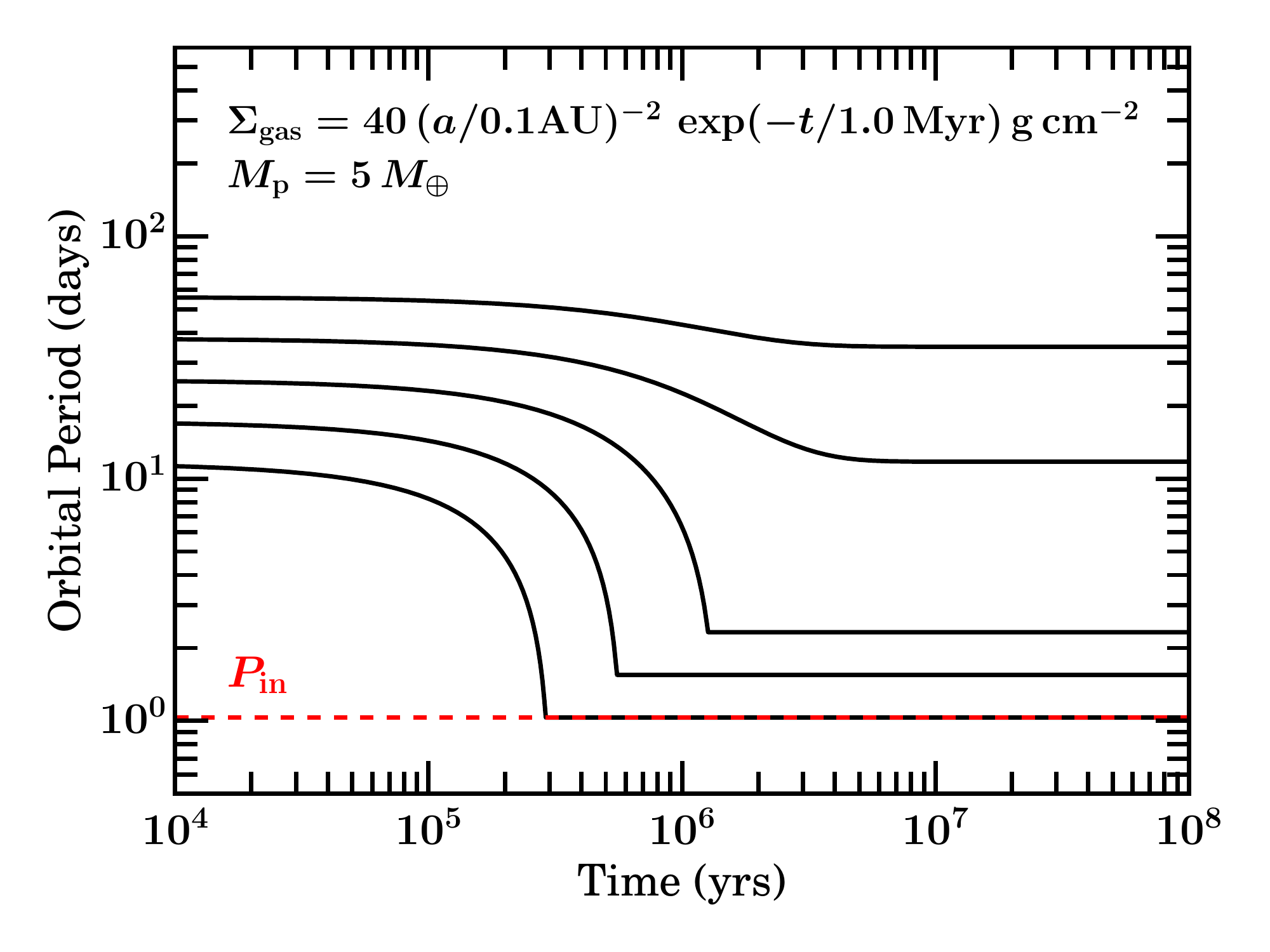}
\caption{How Type I migration might
play out for a
system of sub-Neptunes in a disk that truncates
at period $P_{\rm in} = 1$ day and whose surface
density depletes exponentially with time. We
assume $\beta=2$ and $\gamma=1/2$ for the gas
disk, normalizing the surface density at 0.1 AU
according to the inset equation. At time $t=0$,
$\Sigma_{\rm gas} \sim 10^{-4}\Sigma_{\rm MMSN}$ 
(gas is depleted by $\sim$4 orders of magnitude with respect 
to the solar-composition minimum-mass solar nebula).
All 5 planets have equal mass ($5 M_\oplus$) and
adjacent pairs are initially spaced 14 mutual
Hill radii apart 
(see, e.g., Figure 6 of \citealt{pu15}; note that
in our actual MC models, we randomize the initial
orbital spacings).
Each planet's
trajectory is computed by integrating equation
(\ref{eq1}). The innermost planet is forced to
halt once it reaches the disk's inner edge.
Under the ``resonance lock'' assumption,
planets that follow the first planet are stopped once they
migrate, convergently, into 3:2 resonance.
The final state comprises a chain of 3 resonant 
planets inside 10 days, 1 non-resonant planet 
that has migrated to just outside 10 days (about
a third of its starting period), and 1
outermost planet
that has hardly moved because
its migration timescale
$a/\dot{a}$ exceeded
the disk e-folding time $t_{\rm disk}$, assumed
for this figure to be 1 Myr.
}
    \label{fig3}
\end{figure}

\subsubsection{What happens when planets migrate into the innermost edge}
\label{sec:what}

The first planet to migrate by disk torques 
toward the disk truncation radius will park
in its vicinity---either interior to the disk edge
where the last of the
principal Lindblad torques peters out 
(\citealt{goldreich80}; see also \citealt{lin96});
at the disk edge because of
principal co-rotation torques 
\citep{tanaka02,masset06,terquem07};
or exterior to the disk edge following the reflection
of planet-driven waves off the edge \citep{tsang11}.
In our MC model, we assume for simplicity
that the first planet that can migrate by disk
torques to arrive at the inner edge does so,
and stays there
(until step 3, after which it and
all other planets move further
in by stellar tides).
Experiments that do not park this first planet
at $P = P_{\rm in}$
but instead at $P = P_{\rm in}/2$ (so that the planet is located at
the interior 2:1 resonance with the disk edge) 
change none of our qualitative conclusions.

We have two options for the planets
that subsequently follow this first parked planet.
In one class of MC model (``resonance lock'';
see Figure \ref{fig3} for an illustration), we 
stop a planet when it migrates convergently
into mean-motion resonance with its interior
neighbor. For simplicity, we consider only 3:2
resonances (the choice of 3:2 is motivated by the 
observation that of the small subset of {\it Kepler} planets located
near resonances, most are situated near the 3:2;
\citealt{fabrycky14}).
In a second class of MC model (``merged''),
planets are allowed to migrate unimpeded
to the innermost disk edge and assumed
to merge there with the first planet.
We count the merger product
as a single planet, equivalent to all other planets 
in our final tally of planet occurrence rates 
(this simplification should be
acceptable insofar as observed occurrence rates
based on transit data
are typically tallied by planet radius, which
at fixed bulk density does not change appreciably
when a planet doubles or triples its mass).

The intention behind these two flavors of MC model
is to bracket the range of possible planet-planet
interactions near the disk edge. On the one hand,
convergently migrating planets can become trapped
in a resonant chain \citep[e.g.,][]{terquem07} 
or linger near resonances \citep{deck15}.
Although observed short-period planets are
generally not found in resonance 
\citep[e.g.,][their Figure 8]{sanchis-ojeda14},
orbital instabilities over the Gyr ages of observed
systems can lead to mergers of resonant planets, transforming
them into (lower multiplicity) non-resonant
systems. Numerical N-body integrations
of high-multiplicity systems show that resonances
are preferentially unstable; non-resonant neighbors
appear to destabilize otherwise stable resonant pairs
\citep[][their Figure 5]{mahajan14,pu15}.
Our ``resonance lock'' model captures mergers
of destabilized resonant pairs
insofar as merger products should have the same
period distribution, statistically,
as their progenitors.

On the other hand, planets that collect near the inner
disk edge may never lock into resonances and instead 
collide into each other---hence our ``merged'' model.
Given the potential complexity
of planet-planet and planet-disk interactions
near the disk edge 
\citep[e.g.,][]{tanaka02,masset06,tsang11,deck15}---interactions that
may be modulated by the looming stellar magnetosphere 
\citep[e.g.,][]{romanova16} and/or
disk magnetic fields \citep[e.g.,][]{terquem03}---it would be
unsurprising if planets fail in general
to be captured into resonance in this
messy environment.

\begin{figure*}
    \centering
    \includegraphics[width=\textwidth]{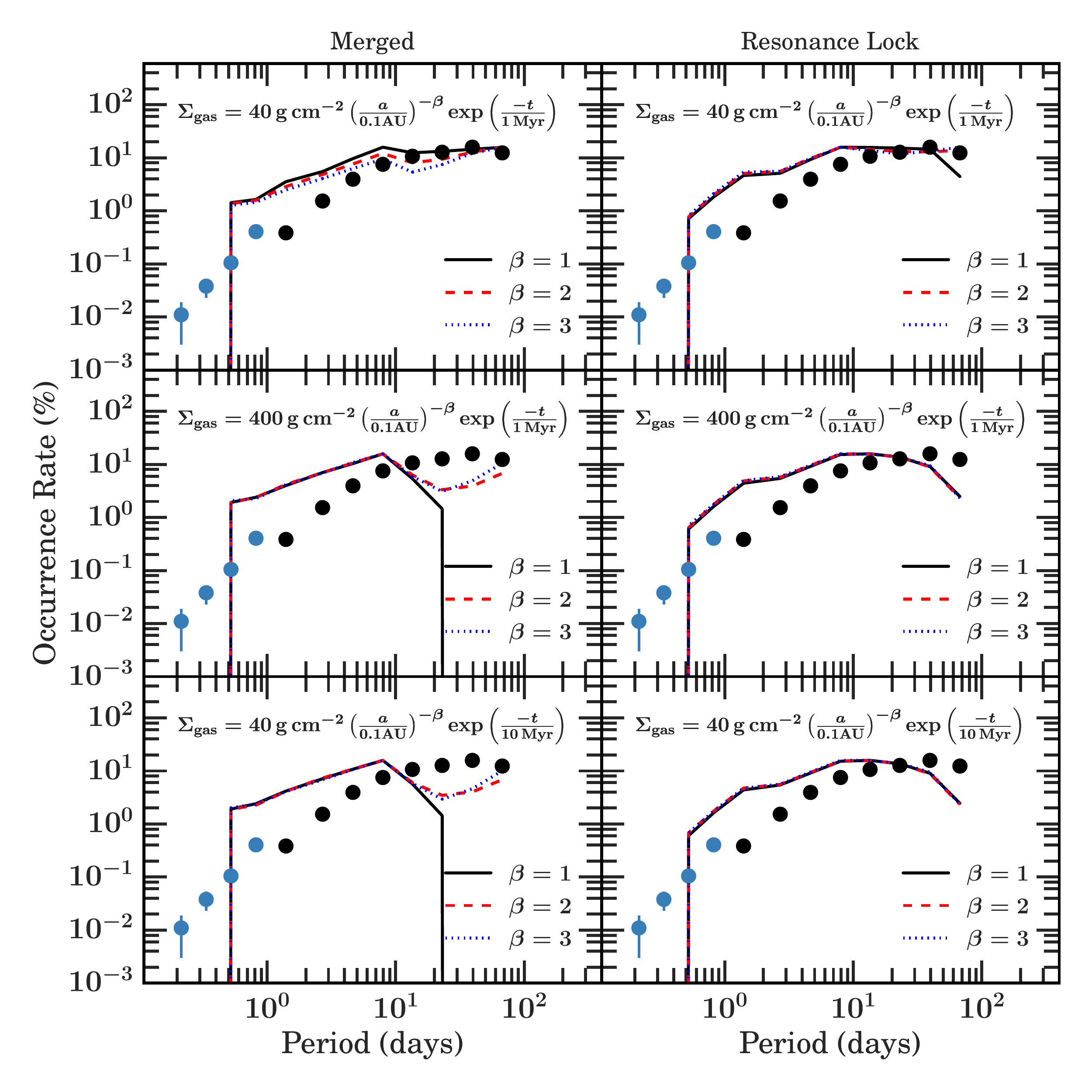}
    \caption{
Model occurrence rates (solid lines) including
Type I disk migration in a dissipating nebula
but excluding tidal migration.
Model curves are computed using the same period bins
as those reported from observations of sub-Neptunes
orbiting FGK stars \citep[][black points]{fressin13}
and GK stars \citep[][blue points]{sanchis-ojeda14},
and normalized such that their maxima
match the maximum observed occurrence rate.    
Left panels correspond to ``merged'' models in which planets
that migrate to the disk inner edge merge there.
Right panels correspond to ``resonance lock'' models
in which planets that migrate convergently into 3:2 resonance
are locked into that resonance.
At $P \lesssim 1$ day,
merged models exhibit a pile-up in the occurrence rate,
while resonance lock models exhibit a sharper decline.
At $P \lesssim 10$ days, models do not differ much when disk
parameters $\beta$, $\Sigma_{\rm gas,0}$, and $t_{\rm disk}$
are varied (see each panel's annotations).
The distribution of disk truncation periods is drawn from
the stellar rotation periods in NGC 2362 (age $\sim 5$ Myr).
The model curves shown here omit tidal migration
and are unable to produce USPs. If we allow disk migration 
to $P = P_{\rm in}/2$ then the models 
have the opposite problem of overproducing USPs.
See later figures
for models that include both disk and tidal
migration.
    }
    \label{fig4}
\end{figure*}

\begin{figure*}
    \centering
    \includegraphics[width=\textwidth]{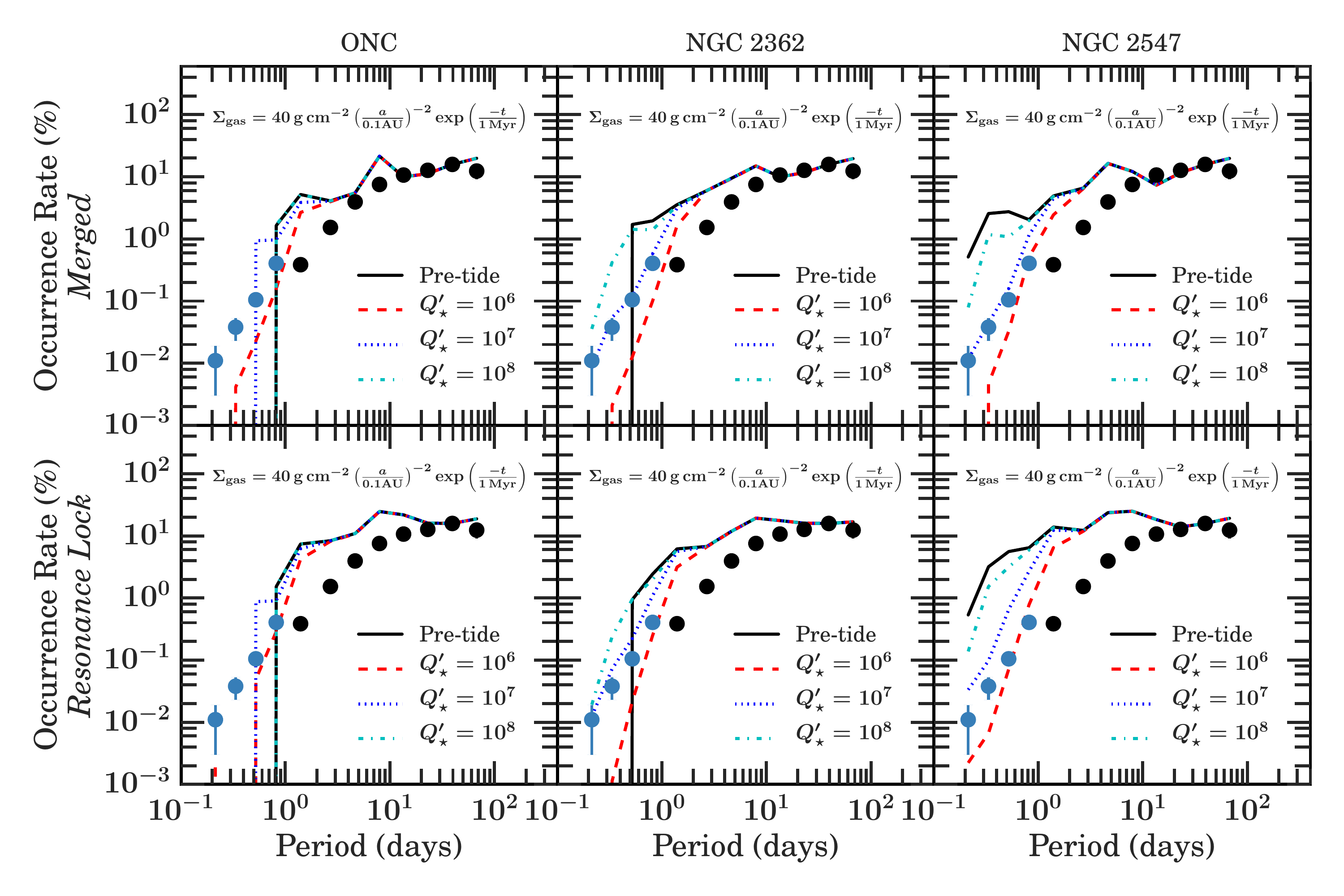}
    \caption{Occurrence rates of planets that have undergone
both disk (Type I) migration in a dissipating nebula, and
tidal migration over 5 Gyr. For reference, black solid lines
include only disk migration and exclude tidal migration.
As in Figure \ref{fig4}, we overplot the observed occurrence rates of
sub-Neptunes around FGK stars with black and blue circles.
All model curves utilize the same period bins as the reported
observations, and are normalized such that they
intersect the observed data point at $P = 39.5$ days.
In each column, disk truncation periods are drawn from 
stellar rotation periods measured for a cluster
of a certain age: 1 Myr for the Orion Nebula Cluster (ONC), 
5 Myr for NGC 2362, and 40 Myr for NGC 2547.
Irrespective of
whether the innermost planets merge at the truncation
radii of their disks (top panels labeled ``merged'') or are
spaced according to mean-motion resonances
(bottom panels labeled ``resonance lock''),
the observed occurrence rate profile for USPs ($P < 1$ day)
can be reproduced by tidal migration,
with best fits corresponding to
a stellar tidal friction parameter $Q'_\star \sim 10^7$ and
an intermediate age (NGC 2362).
Note, however, how in all cases disk migration
overestimates the frequency of planets with
periods between 1 and 10 days.
}
    \label{fig5}
\end{figure*}

\subsection{Step 3: Apply Tidal Orbital Decay}
\label{step3}

The asynchronous equilibrium tide
raised by a planet on its host star
causes its orbital semimajor axis to decay at a rate
\begin{equation}
    \dot{a} = -\frac{9}{2} a^{-11/2} G^{1/2} \frac{M_{\rm p}}{M_\star^{1/2}} \frac{R_\star^5}{Q'_\star}
    \label{eq:tide}
\end{equation}
where $R_\star$ is the stellar radius 
and $Q'_\star$ is the effective tidal 
quality factor (\citealt{goldreich66}; see also
\citealt{ibgui09}).
For every planet, we integrate equation (\ref{eq:tide}) over
a time interval of 5 Gyr to calculate its final
``post-tide'' orbital period.\footnote{
The models shown in this paper
assume that the host star radius is
fixed in time and ignore the pre-main-sequence contraction
phase during which the star is still distended
and the tidal decay rate (which scales as $R_\ast^5$)
is correspondingly enhanced.
We experimented with relaxing this assumption and found
that accounting for the full time
evolution of stellar radius changes
negligibly the final planet occurrence rate profiles.
The pre-main-sequence phase lasts only
$\sim$20 to 60 Myr for 1 to 0.5 $M_\odot$ stars,
and is typically much shorter than the
orbital decay timescales
of planets that do not fall onto the star.}
Equation (\ref{eq:tide}) assumes that 
planets remain on circular orbits, and that
planet orbital periods are shorter than the stellar rotation
period (otherwise planets would migrate outward).
The latter assumption
becomes safer as stars age beyond the 
zero age main sequence and slow their spins.
Although the assumption is violated for the planets with the longest
periods (say $\gtrsim$ 10 days), such planets hardly
migrate by tides anyway because of the steep dependence
of tidal friction on $a$.
We do not consider planet-planet interactions during 
tidal migration as tides act to
separate orbits: 
divergently migrating planets
can cross resonances but do not
lock into them,
and the eccentricities that sub-Neptunes impart to one another
when crossing resonances are small 
\citep[see, e.g.,][]{dermott88}.

For FGK dwarfs, we fix
$R_\star = 1 R_\odot$ and vary $Q'_\star \in \{10^6, 10^7, 10^8\}$,
a range that overlaps with estimates
based on observations and modeling of hot Jupiters
\citep[e.g.,][]{matsumura08,schlaufman10,hansen10,penev12}.
For M dwarfs, we fix $R_\star = 0.5 R_\odot$ and vary 
$Q'_\star \in \{10^4, 10^5, 10^6, 10^7, 10^8\}$. The smaller
values of $Q'_\star$ for M dwarfs are motivated by the
idea that stellar tidal dissipation mostly
occurs in convective zones, which are larger
for lower mass stars \citep[see, e.g.,][their Figure 3]{hansen12}.

Having computed the final periods of all planets
in all disks, we tally them up to calculate model
occurrence rates vs.~orbital period, using the same period bins
as reported by the observations 
(\citealt{fressin13} and \citealt{sanchis-ojeda14}
for FGK dwarfs; 
\citealt{dressing15} for M dwarfs). We construct both
probability distribution functions (PDFs) and corresponding
cumulative distribution functions (PDFs) 
to compare directly with their observational counterparts.
This comparison is done by eye---we are looking 
for broad trends, and aiming to identify only
in an order-of-magnitude sense
those regions of parameter space compatible with the
observations. Our choices on how to normalize our PDFs
and CDFs are given in the figure captions.

\begin{figure*}
    \centering
    \includegraphics[width=\textwidth]{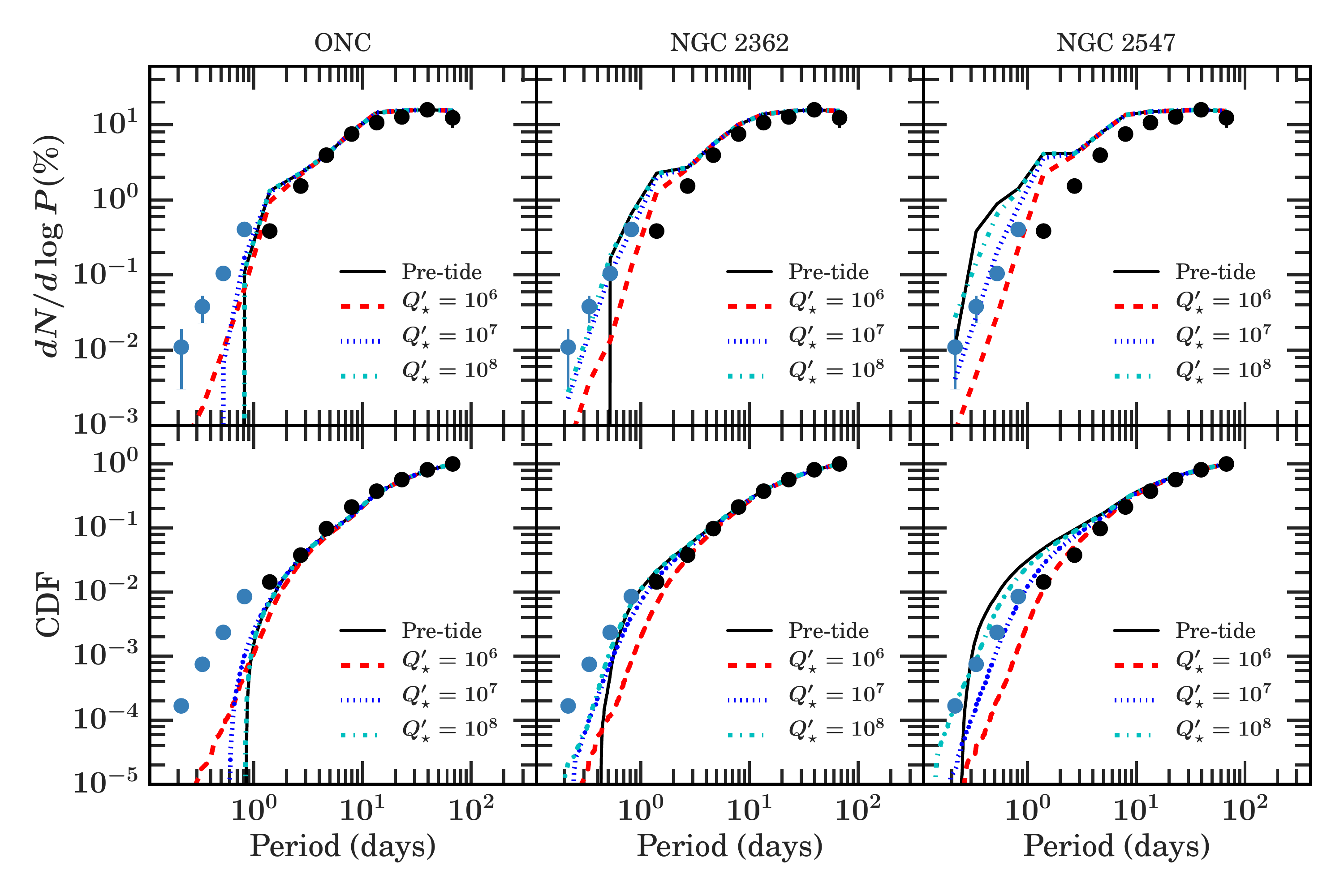}
    \caption{Occurrence rates of planets
that form ``in situ'' (i.e., at locations
randomly and uniformly distributed in $\log P$)
and that subsequently undergo 
tidal migration over 5 Gyr.
The format of this figure is similar to that of Figure \ref{fig5},
except that the bottom row of panels plots the
cumulative distribution function (CDF), 
normalized to unity
at the longest period bin 
(i.e., only counting planets with $P \leq 67.5$ days).
Tidal migration appears necessary to produce USPs at $P < 1$ day.
By eye, best fits correspond to $Q'_\star \sim 10^7$--$10^8$
and disk truncation periods drawn from intermediate-to-late
age clusters like NGC 2362 and NGC 2547.
These in-situ+tidal
models do better than corresponding
disk+tidal migration models (cf.~Figure \ref{fig5}
and see also Figure \ref{fig7}) at
reproducing the number of planets with periods
between 1 and 10 days;
at the same time, they tend to underpredict the number
of USPs at $P \lesssim 0.5$ days.
}
    \label{fig6}
\end{figure*}

\begin{figure*}
    \centering
    \includegraphics[width=\textwidth]{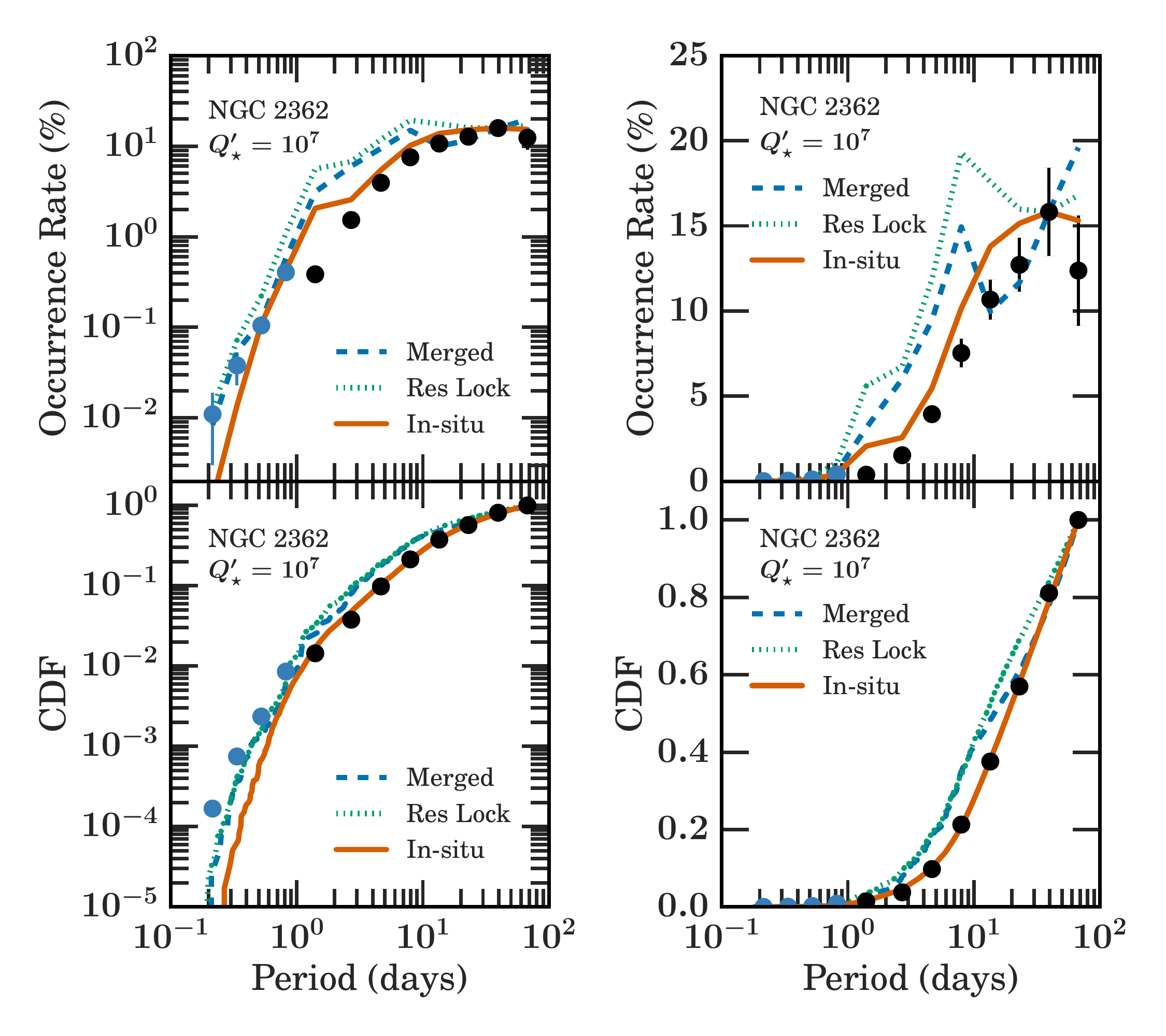}
    \caption{Zooming in on some best-fitting 
disk+tidal and in-situ+tidal models, selected 
from Figures \ref{fig5} and \ref{fig6}, respectively.
In the disk+tidal migration models (blue and cyan curves),
the dissipating nebula is characterized by
$\Sigma_{\rm gas} = 40\,{\rm g \, cm^{-2}} (a / 0.1 \,{\rm AU})^{-2}
\exp (-t / 1 \,{\rm Myr})$. 
The linear scaling shown at right emphasizes that the
disk+tidal migration
models overpredict the number of planets having periods
between 1 and 10 days. In particular, both the
``merged'' and ``resonance lock''
models produce pile-ups just inside 10 days that are not
seen in the observations. In this period range, the in-situ+tidal
model (red curve) better reproduces the data.
But at $P < 1$ day, the situation reverses, as seen in the
logarithmic scaling at left.
Disk migration transports planets efficiently to the
shortest periods where
tides can bring them still further in; thus,
disk+tidal models tend to produce more USPs than
in-situ+tidal models, in better agreement
with observations.
}
    \label{fig7}
\end{figure*}

\subsection{Results for FGK Host Stars}
\label{sec:FGKresults}

We begin by reviewing the results of MC models 
for disk migration, first omitting the
effects of tidal orbital decay in \S\ref{sec:diskFGK1}
and then incorporating them in \S\ref{sec:diskFGK2}.
In-situ + tidal models are presented
in \S\ref{sec:insituFGK}.

\subsubsection{Disk migration without tidal friction}
\label{sec:diskFGK1}

Figure \ref{fig4} shows that despite
the many free parameters,
our disk-migration-only models
give practically the same results:
overprediction of the number
of short-period planets
at $0.5 \, {\rm d} \lesssim P \lesssim 10$ d,
a tendency to underpredict longer-period planets
at $P \gtrsim 10$ d,
and a failure to capture ultra-short-period
planets at $P \lesssim 0.5$ d.
Disk migration scoops out too many planets
at large $P$ to overpopulate small $P$.

That bins at $P < 0.5$ d are empty
follows from $\min P_\star = 0.5$ d
(for NGC 2362, the stellar cluster used to make
Figure \ref{fig4}), and our assumption that
planets do not migrate
past the disk truncation period $P_{\rm in} = P_\star$.
We have tried relaxing this assumption, allowing
planets to migrate to $P_{\rm in}/2$ (i.e., parking them
at 2:1 resonance with the disk edge; see \S\ref{sec:what}).
This succeeds in creating USPs, but merely extends
the problem of overproducing short-period planets
(relative to the observations) to $P < 0.5$ d (data not shown).

The occurrence rate profiles at $P \lesssim 10$ d
appear almost identical over a wide range of disk
parameters ($\beta$, $\Sigma_{\rm gas,0}$, $t_{\rm disk}$)
because ``all roads lead to Rome'':
various disk migration histories
all end with the same outcome of planets converging
onto disk inner edges, whose
period distribution is fixed by the distribution of
stellar rotation periods.
The details of the planet distribution in the vicinity
of the edge depend on what assumption
we adopt: ``merged'' models pile up
planets at $P \sim 0.5$--1 d,
while ``resonance lock'' models 
spread planets out more evenly in $\log P$
from $\sim$1--3 d (under our assumption that planets
lock into 3:2 resonance; in general, for a $j:(j-1)$ resonance,
the pile-up sharpens with increasing $j$).
At longer $P \gtrsim 10$ d,
there is more variation between disk models.
In particular, $\beta = 1$ yields such
high disk densities at large orbital radius that all planets
at long period are flushed inward by disk migration.

\subsubsection{Disk migration with tidal friction}
\label{sec:diskFGK2}

Incorporating tidal inspiral
into our disk migration models
can yield improved fits
to the occurrence rate of USPs
at $P \lesssim 1$ d. 
In Figure \ref{fig5} we apply tidal decay
to one of the disk-migration-only models
shown in Figure \ref{fig4} 
($\Sigma_{\rm gas,0} = 40\,{\rm g\,cm^{-2}}$, $\beta = 2$,
and $t_{\rm disk} = 1$ Myr).
From the grid of models in Figure \ref{fig5},
it appears that using stellar rotation periods from
NGC 2362 and $Q'_\star = 10^7$
works best for reproducing USPs, either
in merged or resonance lock models (middle column, blue curves).

Drawing disk truncation periods from older clusters,
which contain more rapidly rotating stars (Figure \ref{fig2}),
produces more USPs.
In the oldest cluster considered (NGC 2547 at 40 Myr),
typical disk truncation periods are so short that
tidal inspiral is required
to reduce the number of USPs and better match observations.
In the youngest cluster considered (ONC at 1 Myr),
tidal inspiral is also required, but for the opposite
reason: disk truncation periods here are too long and no 
USPs are generated without tidal migration.
The case of NGC 2362 is intermediate and exhibits
both behaviors.

All disk+tidal models, however,
overproduce planets with periods between $\sim$1
and 10 days. This is the same problem as seen in
disk-migration-only models (\S\ref{sec:diskFGK1}).
Tidal orbital decay
has too short a reach to eliminate
this excess.

\subsubsection{In-situ models with tidal friction}
\label{sec:insituFGK}

Figure \ref{fig6} showcases in-situ MC models,
with and without tidal friction. At $P \gtrsim 3$ d,
these models fit the observations remarkably well,
performing better than the disk+tidal migration
models in this period regime. Simply assuming that 
sub-Neptunes are randomly distributed in their disks
(in a uniform $\log P$ sense)
avoids the overproduction-at-short-$P$ /
underproduction-at-long-$P$ problem 
introduced by disk migration.

Ultra-short period planets can be reproduced,
approximately, by tidal orbital decay of planets located 
near the inner edges of their disks.
Drawing the shorter disk truncation periods from
older clusters like NGC 2362 and NGC 2547
works best by giving planets a ``headstart''
toward the shortest periods.

\begin{figure*}
    \centering
    \includegraphics[width=\textwidth]{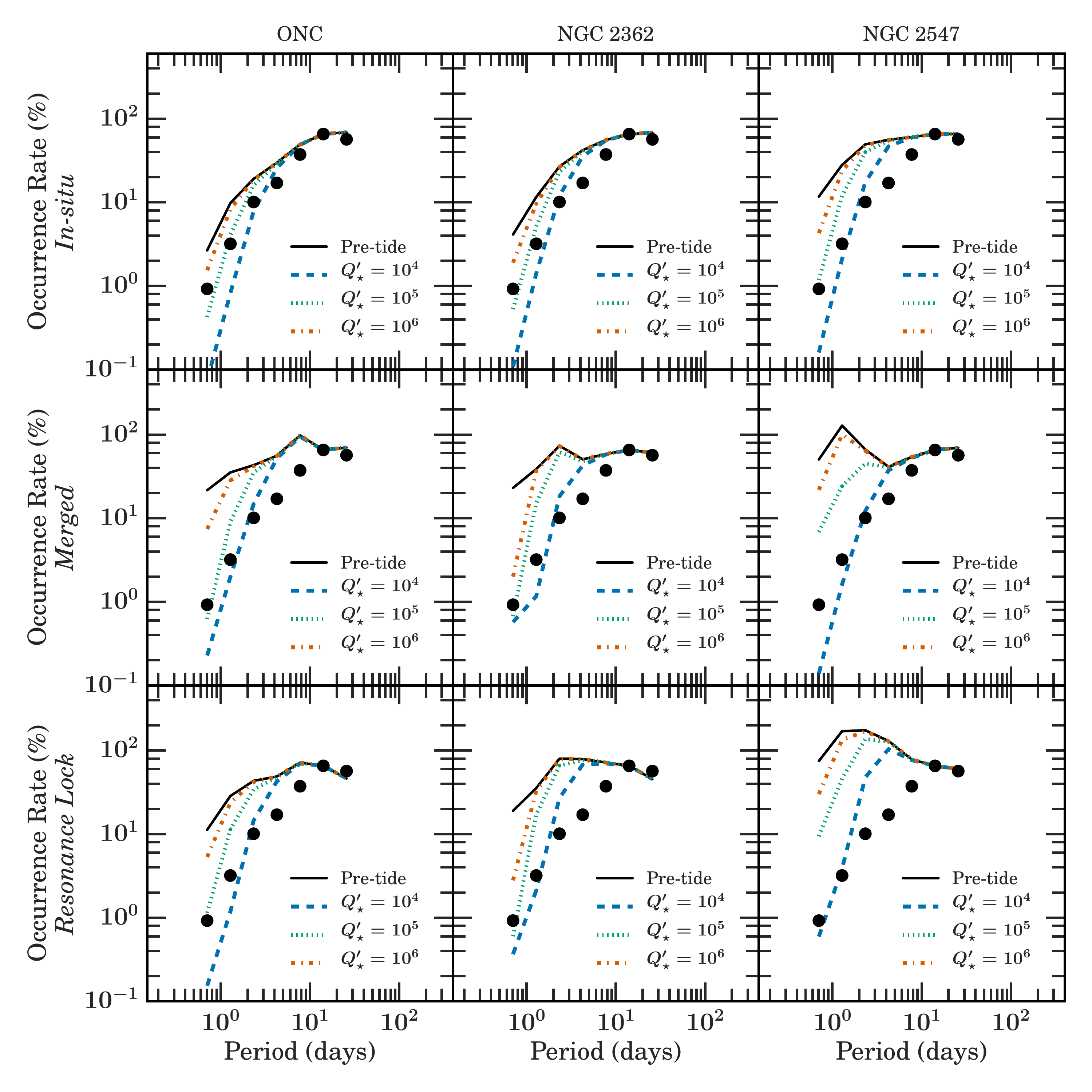}
    \caption{
    Occurrence rate profiles 
    of sub-Neptunes around M dwarfs
    from in-situ+tidal models (top row), 
    disk+tidal models in which planets merge 
    at the disk edge (middle row), and 
    disk+tidal models in which the innermost planets 
    lock into a resonant chain (bottom row). 
    Disk truncation radii are drawn from 
    stellar rotation periods $P_\star$ measured in 
    clusters of varying age, including the
    Orion Nebula Cluster (ONC, 1 Myr, left column), 
    NGC 2362 (5 Myr, middle column),
    and NGC 2547 (40 Myr, right column).
    The rate of orbital decay by stellar tides 
    is calculated using equation (\ref{eq:tide})
    with $R_\star = 0.5 R_\odot$ 
    and $M_\star = 0.5 M_\odot$.
    Observational data (black circles) are adopted from Figure 12 of 
    \citet{dressing15}.
    All model curves are computed using the 
    same period bins as used in the reported 
    observations, and are normalized such that 
    they intersect the observed data point at 14 days.
    The dissipating nebula underlying the disk+tidal models obeys 
$\Sigma_{\rm gas} = 40\,{\rm g \, cm^{-2}} (a / 0.1 \,{\rm AU})^{-1}
\exp (-t / 1 \,{\rm Myr})$. 
    The in-situ+tidal models with $P_\star$ from the ONC 
    and with $Q'_\star \sim 10^5$ agree best with the 
    observations.   
    }
    \label{fig8}
\end{figure*}

Figure \ref{fig7} zooms in to directly compare
one of the better fitting in-situ+tide models
with its disk+tide counterpart.
With respect to the observations,
the in-situ+tide model does better
than the disk+tide model at $P \sim 1$--10 days
by not overproducing the number of planets.
But for that same reason, the disk+tide model
does better than the in-situ+tide model
at $P \lesssim 1$ day: disk migration generates
more USPs by supplying more planets for tidal 
inspiral, and matches the observations at 
ultra-short periods more closely.
Both classes of model overpredict the number of planets 
at $P \sim$1--3 days.

\begin{figure*}
    \centering
    \includegraphics[width=\textwidth]{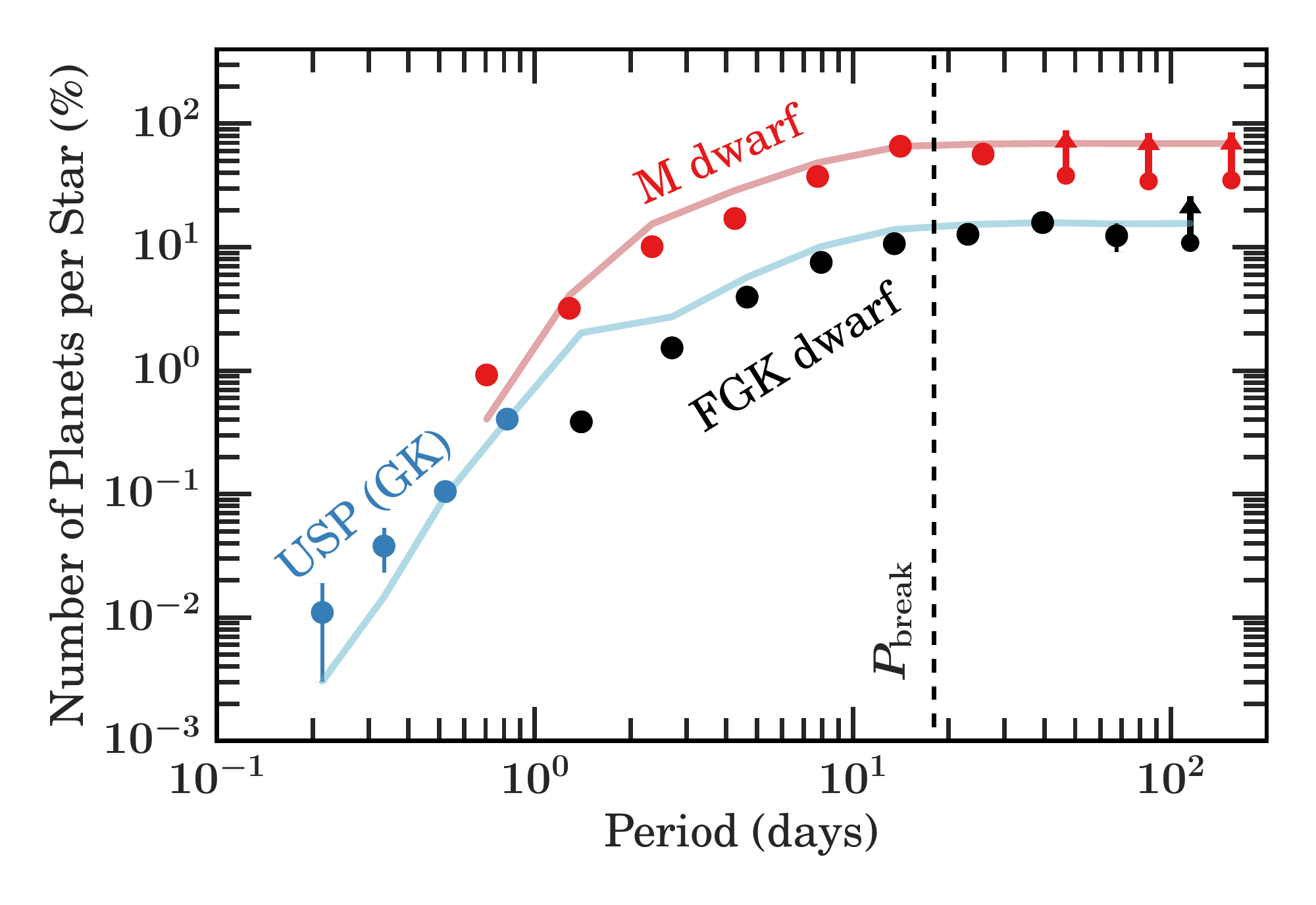}
    \caption{In-situ planet formation in disks truncated at co-rotation by host star magnetospheres, in combination with tidal migration, can reproduce the observed occurrence rates of sub-Neptunes. Black and blue points represent observations of planets orbiting FGK dwarfs by \citet{fressin13} and \citet{sanchis-ojeda14}, respectively, and red points represent observations of planets orbiting early M dwarfs by \citet{dressing15}. 
    Points without arrows correspond
    to sub-Neptunes larger than $0.5 R_\oplus$
    for M dwarf hosts, and larger than
    $0.8 R_\oplus$ for FGK hosts.
    Points with arrows represent 
    sub-Neptunes larger than $1 R_\oplus$ (M dwarfs),
    and larger than $1.25 R_\oplus$
    (FGK dwarfs).
    Our best-fitting model for FGK hosts uses stellar rotation periods drawn from NGC 2362 at an age of 5 Myr, and $Q'_\star = 10^7$. Our best-fitting model for M-type hosts uses rotation periods drawn from the Orion Nebula Cluster at an age of 1 Myr, and $Q'_\star = 10^5$. These same models are plotted in Figures \ref{fig6}, \ref{fig7}, and \ref{fig8}, but we display them again here for ease of viewing.}
    \label{fig9new}
\end{figure*}

\subsection{Results for M Host Stars}
\label{sec:Mresults}

Figure \ref{fig8} compares 
model occurrence rates for 
planets orbiting M dwarfs against observations. 
The best agreement is obtained for
in-situ+tidal models calculated by 
drawing $P_{\rm in}$ from the ONC 
and by adopting $Q'_\star \sim 10^5$.
Using these same assumptions
in disk-induced migration models gives worse fits 
characterized by an excess of planets 
inside $\sim$10 days.
Appealing to stronger tidal dissipation can 
remove much of this excess but
simultaneously destroys ultra-short period planets 
by shuttling them to the stellar surface.

We require a significantly smaller $Q'_\star$ to 
produce a best-fitting in-situ+tidal model for M dwarfs 
($Q'_\star \sim 10^5$) as compared to FGK dwarfs ($Q'_\star \sim 10^7$).
M stars tend to rotate faster than FGK stars,
so stronger tides 
are needed to remove the excess of
planets near $\sim$1--2 days.
The smaller $Q'_\star$ for M dwarfs
supports the idea that the bulk of 
stellar tidal dissipation occurs in convective zones, 
which are larger for lower mass stars
\citep[e.g.,][and references therein]{hansen12}. 

Drawing $P_{\rm in}$ from
clusters older than the ONC
overestimates the number of planets 
inside 10 days, as stars spin up with age
and truncate their disks at shorter periods.
This effect is amplified 
for M stars over FGK stars
because M dwarfs spin up more
rapidly as they get older;
see in Figure \ref{fig2} how the peak 
of the period distribution shifts 
more dramatically with age for lower mass stars.
The excess population is especially 
difficult to remove when planets enter into and remain 
in a ``resonance lock'' near the disk edge.
Resonance lock models distribute planets broadly
over this period range, beyond the reach of tidal friction
in M stars even for $Q'_\star = 10^4$.

\section{Summary and Discussion}
\label{sec:discuss}

Around FGK and M stars,
sub-Neptunes appear evenly distributed
at orbital periods $P \gtrsim 10$ days,
but become increasingly rare
toward shorter periods.
Through Monte Carlo population synthesis calculations,
we demonstrated how this orbital architecture
arises from disks truncated at co-rotation
with their host star magnetospheres,
and from orbital decay
driven by stellar tidal dissipation.
The drop-off in planet 
occurrence rate at $P \sim 1$--10 days 
traces the distribution of
planets near the inner edges of their parent disks;
in turn, the distribution of disk edges 
reflects the distribution of stellar rotation periods 
in young clusters 1--40 Myr old.
At the same time, despite disk truncation,
planets can migrate inward to ``ultra-short'' periods ($P < 1$ day)
by the asynchronous equilibrium tides they raise on their
host stars.

We found better fits to observed occurrence rates when 
planets were randomly distributed in log $P$ 
within their parent gas disks---as
might be expected if they formed in situ---rather than
brought to disk edges by disk torques.
Figure \ref{fig9new}
recaps how well our in-situ+tidal models reproduce the occurrence rates
of planets orbiting FGK dwarfs and early M dwarfs.
However, none of our models, predicated either on
in-situ formation or disk migration,
fits the data perfectly; the former do better 
at large $P$ while the latter do better at short $P$.
In-situ models might yield improved fits 
if we had a more accurate theory for tidal orbital decay,
or if the distribution of 
disk inner edges were broader
than we have assumed.
Such broadening could be achieved by relaxing
our assumption that disks truncate exactly at co-rotation,
or by issuing sub-Neptunes
from disks with a range of dissipation timescales.
For example, the disk truncation period
could be drawn from a mix of
stellar rotation periods characterizing
both the young Orion Nebula Cluster
and the older NGC 2547 (but see section \ref{ssec:stellarmass}).
It is also possible that the stellar rotation period
distributions we have used are not entirely representative.
The cluster Cep OB3b has a similar age (4--5 Myr) as NGC 2362,
but its median rotation period for M dwarfs may be
$\sim$1.6 times longer, assuming its
low-mass stars are correctly identified as such
\citep{littlefair10}.

Below we explore how our theory 
might be extended to bear on
other observed properties of 
short-period sub-Neptunes
and their host stars, 
attempting where possible to make predictions 
for future observations.

\begin{figure}
    \centering
    \includegraphics[width=0.5\textwidth]{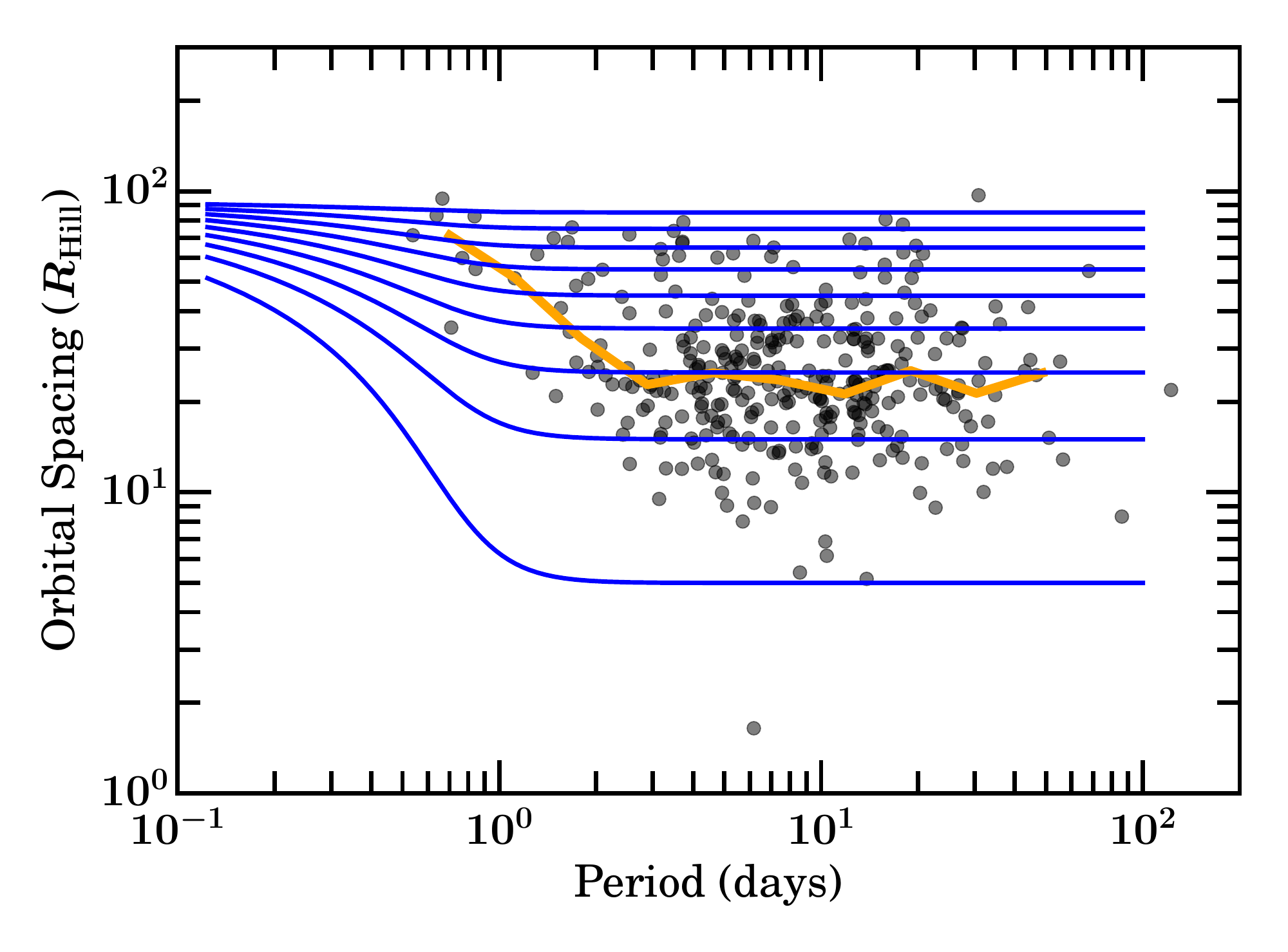}
    \caption{Observational evidence for how
stellar tidal friction widens separations
between planets most effectively at the shortest
orbital periods.
Each gray circle represents the orbital spacing,
measured in mutual Hill radii $R_{\rm Hill}$,
between the two shortest-period transiting planets
(``tranets'') in a given multi-tranet system, and
is plotted against the shorter period of the pair.
These data, downloaded on 2016 Dec 5
from the NASA Exoplanet Archive
and based on Quarters 1--16 of the {\it Kepler} 
space mission, include only sub-Neptunes ($R < 4 R_\oplus$).
In calculating mutual Hill radii, we assumed all tranets
to have $M_{\rm p} = 5 M_\oplus$.
The median observed spacing, plotted as a thick
orange solid line, is fairly constant at $P > 2$ days
but grows dramatically toward shorter periods.
Such behavior is expected, qualitatively,
from tidal migration.
Blue lines indicate theoretical orbital spacings
between adjacent
pairs of $5 M_\oplus$ planets after 5 Gyrs of
tidal orbital decay; the central star on which tides are raised
is assumed to have
mass $M_\star = M_\odot$, radius $R_\star = R_\odot$,
and tidal quality factor $Q'_\star = 10^7$.
From bottom to top,
theory curves correspond to pairs
of planets separated initially (before tides act)
by 5, 15, \ldots, 95 $R_{\rm Hill}$, and each
is plotted against the final (post-tide) orbital period
of the inner member of a pair.
}
    \label{fig9}
\end{figure}

\subsection{Orbital Spacings of USPs}

We have shown in this paper the essential role tides play 
in creating ultra-short period planets (USPs).
Another line of evidence pointing to the influence of tides 
can be found by examining planetary orbital spacings.
Because tidal orbital decay generally proceeds
more quickly at smaller periods,
interplanetary spacings should be stretched
out more (in a fractional sense)
at small $P$ than at large $P$.
Just such a signature is observed:
USPs are more widely spaced than their longer period counterparts
\citep{steffen13}.
In Figure \ref{fig9} we demonstrate how orbital spacings
widen preferentially at the shortest periods
because of tides.
Comparison of our model with observations shows 
only qualitative agreement;
the empirical data (gray circles and orange line)
suggest that tidal friction in reality is more effective
at drawing planets apart than our constant-$Q'_\star$
theory allows (blue curves).
A better theory for tidal friction, one incorporating
dynamical tides and the evolution of stellar spin and structure,
is needed to better reproduce the data
\citep[see, e.g.,][and references therein]{bolmont16}.

\begin{figure}
    \centering
    \includegraphics[width=0.5\textwidth]{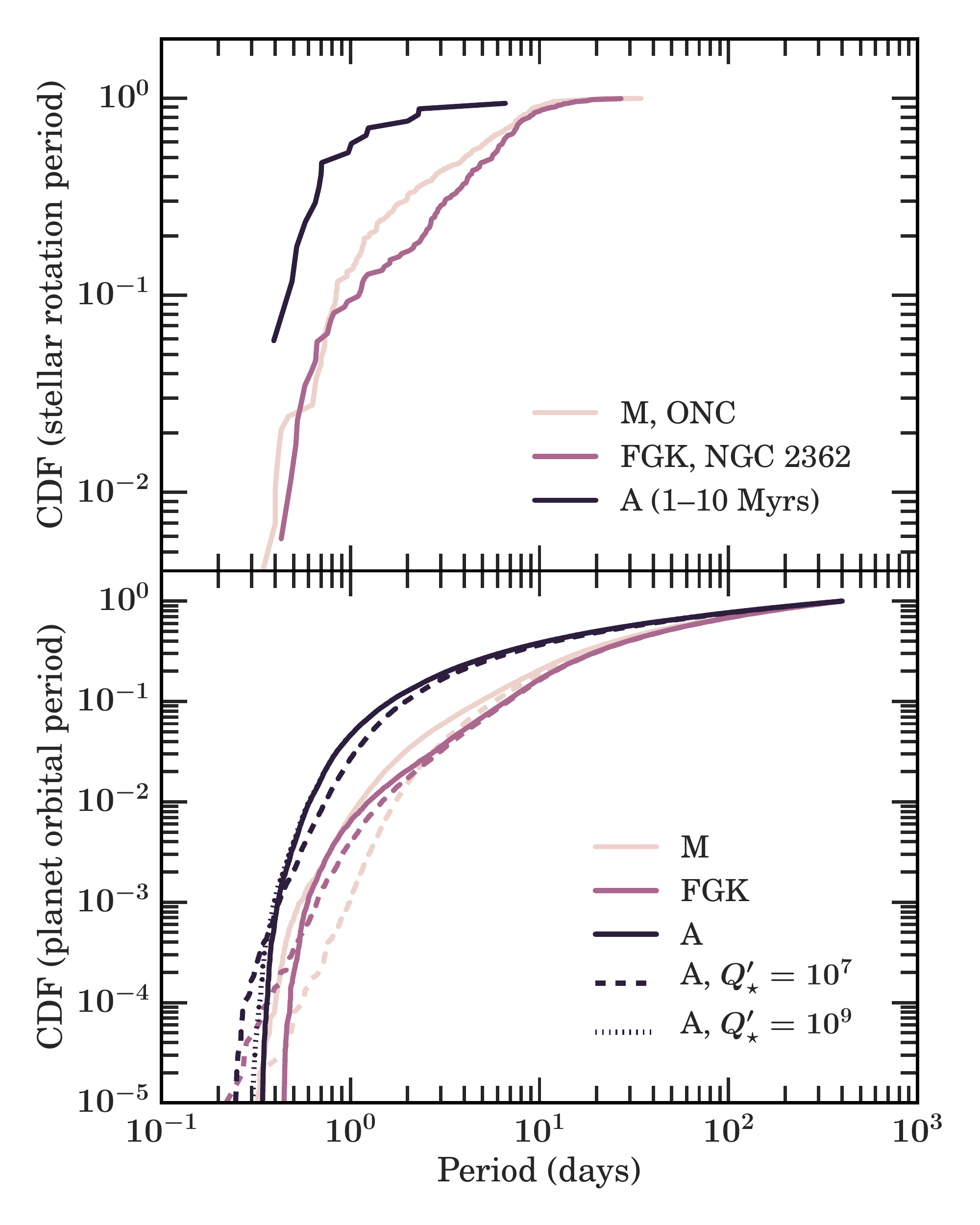}
    \caption{
    Pre-main sequence A stars rotate faster than 
    pre-main sequence FGK or M stars (top panel), 
    and we therefore expect that the planet occurrence rate for A stars 
    breaks at a shorter period $P_{\rm break}$ than for FGKM stars (bottom panel).
    Rotation periods of 1--10 Myr old Herbig Ae stars 
    are taken from Table 7 of \citet{hubrig09}, 
    updated using Table 2 of \citet{alecian13}.
    These rotation periods were calculated from spectroscopically determined 
    stellar radii and rotation velocities  
    de-projected using disk inclinations; the latter were measured either from resolved disk images or inferred from the line profile shapes of the Mg II or H$\alpha$ emission lines.
    Pre-tide and post-tide models, all assuming in-situ formation,
    are shown in the bottom panel as solid and dashed lines, respectively;
    for FGKM stars, we use the model parameters
    that yield the best agreement with observations
    (ONC and $Q'_\star = 10^5$ for M;
    NGC 2362 and $Q'_\star = 10^7$ for FGK).
    To account for the shorter main sequence lifetime of higher 
    mass stars, we apply tidal migration around A stars
    for only 1 Gyr (vs.~5 Gyrs for FGKM stars).
    For planets around A stars,
    within a plausible range of $Q'_\star$
    (and taking $M_\star = 2M_\odot$ and $R_\star = 2R_\odot$),
    we find that $P_{\rm break} \sim 1$ day,
    considerably shorter
    than the period break for FGKM stars ($\sim$10 days).
    There is a weaker secondary break at $\sim$1 day 
    around FGK dwarfs that reflects the bimodal distribution 
    of rotation periods for stars in 
    this mass range (Figure \ref{fig2};
    see also \citealt{bouvier07}
    and references therein).
    }
    \label{fig10}
\end{figure}

\subsection{Trends with Stellar Mass}
\label{ssec:stellarmass}

\citet{mulders15} 
point out how the sub-Neptune occurrence rate 
breaks at $P_{\rm break} \sim 10$ days 
regardless of whether the host star
is of spectral type F, G, K, or M.
In our theory, the break and drop-off at shorter periods
reflect the distribution of stellar spin
periods $P_\star$. 
We can reproduce the invariance of $P_{\rm break}$ 
by drawing $P_\star$ from a distribution that does not
vary with stellar mass.
The Orion Nebula Cluster (ONC) provides such a
spin distribution,
which peaks at $\sim$10 days
irrespective of whether the stars are $\sim$1 $M_\odot$ or $\sim$0.3 $M_\odot$
(compare top and bottom panels of Figure \ref{fig2}).
Drawing $P_\star$ from the ONC implies that
planets emerge from disks that dissipate
over the ONC age of $\sim$1 Myr.
One problem with our ONC-based models is that they
underestimate the number of USPs,
particularly around FGK stars 
(left column of Figure \ref{fig6});
not enough stars rotate with periods
faster than a day.
Ultra-short period planets
are produced more easily around 
older, more rapidly rotating stars
(middle and right columns of Figure \ref{fig6}),
but these have $P_\star$ distributions
that do vary with stellar mass (Figure \ref{fig2})
and therefore have trouble reproducing
the invariance of $P_{\rm break}$ (note how
the model break period in the middle and the right columns 
of Figure \ref{fig8} misses the observed break period by a factor of 5).
Simultaneously fitting both USPs and their longer period 
counterparts near the break is a challenge,
requiring improvements in our understanding of disk truncation and tides.

There is also room for improvement for observations:
the USP occurrence rates from \citet{sanchis-ojeda14}
appear systematically offset from the \citet{fressin13} data 
at longer periods; see how the blue points
in Figure \ref{fig9new} appear to be shifted higher
than the black points.  
Whether this offset is real or an artifact of combining two different datasets is unclear; removal of the offset would ameliorate if not solve the problem noted above with our ONC-based models.
A uniform analysis
is needed to measure planet occurrence rates from $\sim$0.1 to $> 10$
days, not just for FGK stars but also for M stars.

Detecting and characterizing planets around 
main sequence
A stars is the next frontier. 
The invariance of $P_{\rm break}$ with respect to stellar spectral type 
may not extend to A stars.
Compared to T Tauri stars,
Herbig Ae stars are observed to be
faster
rotators, with typical spin periods of $\sim$1 day
\citep{hubrig09,alecian13}. 
It follows that their gas
disks should truncate at shorter periods.\footnote{Herbig
Ae stars can have magnetospheres.
They are observed to have typical
surface magnetic field strengths 
of $\sim$100 G \citep[see, e.g.,][their Figure 10]{hubrig09}. 
For disk accretion rates of $10^{-7}$--$10^{-10} M_\odot\,{\rm yr}^{-1}$ 
around a star of mass $2\,M_\odot$ and
radius $4\,R_\odot$, magnetospheric truncation radii range between
$\sim$1 and 4 stellar radii.} 
We therefore predict
the occurrence rate of sub-Neptunes 
to break at a correspondingly short period $P_{\rm break} \sim 1$ day
around A stars, as
compared to $P_{\rm break} \sim 10$ days for FGKM stars.
This shift in $P_{\rm break}$ between high and low mass
stars is illustrated in Figure \ref{fig10}.
The break around A stars might occur at even shorter
sub-day periods,
as our spin period data for A stars are taken from
stellar magnetic field studies whose samples are biased
to include more slowly rotating stars \citep{hubrig09,alecian13}.
Figure \ref{fig10} also shows that tidal inspiral
does not much alter
the period break for A stars
for $Q'_\star = 10^7$--$10^9$.
These larger values of $Q'_\star$ 
are motivated by main sequence A stars
lacking outer convective envelopes 
\citep[see, e.g.,][their Figure 4]{mathis15}.
A caveat to our prediction that sub-Neptunes around A stars
exhibit a shorter $P_{\rm break}$ is that we have so far discussed
only disks of gas,
not disks of solids---and sub-Neptunes are primarily composed
of solids.
Dust disks around A stars may not extend 
down to periods of $\sim$1 day because of sublimation
\citep[e.g.,][their Figure 7]{dullemond10}.
Nevertheless, larger solids (planetesimals orders
of magnitude larger than dust grains) are more resistant
to vaporization and can drift, aerodynamically or otherwise,
through gas disks towards their inner edges, 
assembling into planets there.

\acknowledgements
We thank Kevin Covey, Trevor David, 
Dan Fabrycky, Brad Hansen, Brian Jackson, 
Dong Lai, Jing Luan, Gijs Mulders,
Ruth Murray-Clay, James Owen, Ilaria Pascucci, 
Roman Rafikov, Saul Rappaport, Fred Rasio, Hilke Schlichting, Jason Steffen, Scott Tremaine, and Josh Winn for helpful discussions. An anonymous referee provided an exceptionally thoughtful and encouraging report.
This research has made use of the NASA Exoplanet Archive, which is operated by the California Institute of Technology, under contract with the National Aeronautics and Space Administration under the Exoplanet Exploration Program.
EJL was supported in part by the Berkeley Fellowship 
and by NSERC of Canada through PGS D3.
EC acknowledges support from NASA and NSF.

\appendix

\section{Migration in Gas-poor Disks}
\label{sec:app1}

In our MC models with disk-induced
Type I migration, 
we chose gas surface densities so low 
that the planet mass $M_{\rm p}$
exceeds, by roughly two orders of magnitude, the local disk mass 
$\Sigma_{\rm gas}a^2$. For the planet to migrate inward,
it must lose its angular momentum to the disk. 
Because $M_{\rm p} \gg \Sigma_{\rm gas}a^2$,
one might worry that the need to transfer
so much angular momentum
from the planet to the local disk would severely perturb
the latter, so much so that the planet would
cease to undergo normal Type I migration.
Perhaps the planet stops migrating altogether;
or perhaps it opens a deep and wide
gap and migrates at the Type II rate set
by disk viscosity.

Even when $M_{\rm p} \gg \Sigma_{\rm gas}a^2$,
Type I migration might still prevail in disks with large enough viscous
stresses to prevent the opening of gaps and
to transport angular momentum---both the planet's
and the disk's---outward to distances of a few AU.\footnote{By contrast,
in inviscid disks, the inability of the disk to carry away the planet's angular
momentum presents a serious problem for Type I migration.
A ``migration feedback'' torque can stall planets in 
low-mass, laminar disks (\citealt{ward89};
\citealt{ward97}; \citealt{rafikov02}; \citealt{li09},
\citealt{yu10}; \citealt{fung17}).} 
For our chosen disk parameters, we find that as long as the Shakura-Sunyaev
viscosity parameter $\alpha \gtrsim 2\times 10^{-4}$, super-Earths of mass
$5M_\oplus$ will not open gaps at short periods 
\citep{duffell13,fung14}. 
Furthermore, for the disk depletion time to be as long as $t_{\rm disk} \sim 1$
Myr, the disk must be able to transport mass and momentum across large distances
extending well beyond the innermost regions where short-period planets reside. 
For example, assuming $\alpha \gtrsim 2 \times 10^{-4}$ (as per above), the
viscous drainage time of the disk,
$t_{\rm visc} \sim a^2/\nu \sim \Omega a^2 \mu m_{\rm H}/ (\alpha k T)$
(for viscosity $\nu$), is as long as $t_{\rm disk} \sim$ 1 Myr only for
$a \gtrsim 2$ AU. 
The mechanism of disk transport remains unclear, but the magneto-rotational
instability and magnetized winds are perennial candidates, especially in the hot
and well-ionized innermost disk, and perhaps also beyond under gas-poor
conditions \citep[see, e.g.,][]{wang16}.

Even if migration is not of Type I,
its exact form should have little bearing
on our disk+tide migration models
at orbital periods shorter than $\sim$20 days.
As we demonstrated in \S\ref{sec:diskFGK1},
the period distribution
of close-in planets is determined largely
by the distribution of stellar rotation periods,
which sets disk truncation periods.
In our disk+tide models,
it does not much matter how planets are transported 
to disk edges, as long as they arrive there.
We reiterate that our in-situ+tide models,
which omit disk migration altogether,
give overall better fits to the observations,
especially at $P > 1$ day.

\bibliography{usp}
\end{document}